%% file: main.tex
\PassOptionsToPackage{hyperfootnotes=false}{hyperref}
\documentclass[sigconf, nonacm]{acmart}

\AtBeginDocument{%
  }

\usepackage[utf8]{inputenc}%
\usepackage[T1]{fontenc}%
\usepackage{hyperref}%
\usepackage{url}%
\usepackage{booktabs}%
\usepackage{amsfonts}%
\usepackage{nicefrac}%
\usepackage{microtype}%
\usepackage{xcolor}%
\usepackage{xspace}
\usepackage{listings}
\usepackage{caption}
\usepackage{subcaption}
\usepackage{float}
\usepackage{graphicx}
\usepackage{colortbl}
\usepackage{xcolor}

\newcommand{\tool}{WasmMend\xspace}
\newcommand{\DS}{Deep\-Seek-V4.1-Flash\xspace}
\newcommand{\Qwen}{Qwen3.8-Flash\xspace}
\newcommand{\Minimax}{Mini\-Max-M3\xspace}
\newcommand{\Glimmer}{Muse-Glimmer-30B\xspace}
\newcommand{\Mimo}{MiMo-V2.6-Flash\xspace}

\definecolor{bg}{RGB}{248,248,248}
\definecolor{kw}{RGB}{127,0,85}
\definecolor{cm}{RGB}{63,127,95}
\definecolor{gr}{RGB}{140,140,140}
\definecolor{add}{RGB}{0,112,0}
\definecolor{del}{RGB}{180,0,0}
\definecolor{rulecolor}{RGB}{200,200,200}
\definecolor{shellbg}{RGB}{253,246,227}
\definecolor{shellfg}{RGB}{101,123,131}
\definecolor{shellprompt}{RGB}{0,100,0}
\definecolor{shellcmd}{RGB}{7,54,66}
\definecolor{shellerr}{RGB}{220,50,47}
\definecolor{hldel}{RGB}{180,140,0}
\definecolor{hladd}{RGB}{0,90,170}

\usepackage{xcolor}%

\lstdefinestyle{base}{
  basicstyle=\ttfamily\scriptsize,
  backgroundcolor=\color{bg},
  frame=single, rulecolor=\color{rulecolor}, framerule=0.3pt,
  breaklines=true, tabsize=2, showstringspaces=false,
  aboveskip=4pt, belowskip=2pt,
  xleftmargin=3pt, xrightmargin=3pt,
  framexleftmargin=1pt, framexrightmargin=1pt,
}

\lstdefinestyle{code}{
  style=base, language=C,
  keywordstyle=\color{kw}\bfseries,
  commentstyle=\color{cm}\itshape,
  numbers=left, numberstyle=\tiny\color{gr}, numbersep=4pt,
  moredelim=[is][\color{hldel}\bfseries]{«D}{D»},
  moredelim=[is][\color{hladd}\bfseries]{«A}{A»},
  postbreak=\raisebox{0ex}[0ex][0ex]{\ensuremath{\hookrightarrow\space}},
}

\lstdefinestyle{shell}{
  basicstyle=\ttfamily\scriptsize\color{shellfg},
  backgroundcolor=\color{shellbg},
  frame=single, rulecolor=\color{shellbg}, framerule=0.3pt,
  breaklines=true, tabsize=2, showstringspaces=false,
  aboveskip=4pt, belowskip=2pt,
  xleftmargin=4pt, xrightmargin=4pt,
  framexleftmargin=2pt, framexrightmargin=2pt,
  numbers=none,
  moredelim=[is][\color{shellprompt}\bfseries]{`P}{P`},
  moredelim=[is][\color{shellcmd}]{`C}{C`},
  moredelim=[is][\color{shellerr}]{`E}{E`},
}

\usepackage{enumitem}
\usepackage{algpseudocode}
\usepackage{amsmath}
\usepackage{xcolor}
\definecolor{CommentColor}{rgb}{0.1, 0.5, 0.1}

\algrenewcommand{\algorithmiccomment}[1]{\hfill\textcolor{CommentColor}{\textit{// #1}}}

\usepackage{wrapfig}
\usepackage{multirow}
\usepackage[table]{xcolor}

\usepackage[ruled,vlined]{algorithm2e}
\definecolor{deepgreen}{RGB}{0, 128, 0}
\usepackage{setspace}
\title{Reasoning from Traces: Divergence-Guided Agentic Repair of WebAssembly Discrepancies}

\author{Liyan Huang}
\authornote{Equal Contribution.}
\email{liyanhua@usc.edu}
\affiliation{%
  \institution{University of Southern California}
  \city{Los Angeles}
  \country{United States}
}

\author{Kaicheng Wang}
\authornotemark[1]
\email{wangkaic@usc.edu}
\affiliation{%
  \institution{University of Southern California}
  \city{Los Angeles}
  \country{United States}
}

\author{Amy Zhang}
\email{amyz@usc.edu}
\affiliation{%
  \institution{University of Southern California}
  \city{Los Angeles}
  \country{United States}
}

\author{Weihang Wang}
\email{weihangw@usc.edu}
\affiliation{%
  \institution{University of Southern California}
  \city{Los Angeles}
  \country{United States}
}

\begin{document}

\input{tex/Abstract}



\maketitle

\input{tex/Introduction}
\input{tex/RelatedWork}

\input{tex/Methodology/Methodology}

\input{tex/Evaluation/Evaluation}
\input{tex/Appendix/ThreatsToValidity}
\input{tex/Conclusion}

\input{main-bibliography.tex}
\input{tex/Appendix/Appendix}

\clearpage
\end{document}

%% file: tex/Abstract.tex
\begin{abstract}

WebAssembly (Wasm) promises seamless reuse of C/C++ codebases as portable, efficient, sandboxed binaries for execution on the Web. In practice, however, this promise often falls short: recent studies show that cross-compiling C/C++ source to Wasm and native binaries frequently produces runtime discrepancies due to differences in library implementation and compiler bugs. Since the root causes are platform-level and hidden beneath the source code, even state-of-the-art LLM-based coding agents can struggle to diagnose and repair these discrepancies.
In this paper, we present WasmMend, the first system for automatically repairing Native-Wasm functional discrepancies. WasmMend transforms open-ended debugging into a two-stage process. First, a novel differential trace analysis approach localizes the function where Wasm and native executions first diverge. Second, guided by this localization, a multi-agent system coordinates separate agents to interpret platform-specific behavior and synthesizes patch plans. Experiments on discrepancies from real-world C/C++ projects show that, across five backbone models, WasmMend achieves a fix rate of up to 84.8\%, with average gains of at least 37.2 percentage points over the evaluated open-source coding agent baselines, while maintaining comparable cost.
\end{abstract}

%% file: tex/Introduction.tex
\section{Introduction}
\label{sec:intro}

WebAssembly (Wasm) is a Web standard that specifies a portable binary instruction format. It serves as a compilation target for high-level languages such as C/C++, enabling secure and fast execution on the Web~\cite{bringing_up_to_speed_wasm}.  
A key promise of Wasm is seamless code migration: mature cross-compilation toolchains~\cite{cheerp, Emscripten} allow developers to reuse existing C/C++ codebases as Wasm binaries with minimal effort~\cite{WasmDiff}. In the real world, 64.2\% of Wasm modules originate from C/C++~\cite{WasmBench}, highlighting the need for reliable migration.

However, recent studies~\cite{WasmChecker, PC2W, WasmDiff} reveal that the migration promise often breaks in practice. 
Many platform-level discrepancies, such as varied library implementations and compiler bugs, can yield executable native (e.g., x86) and Wasm binaries that silently diverge in behavior or crash.
For example, standard POSIX operations can trigger crashes in Wasm. Figure~\ref{fig:mmap-bug} illustrates a discrepancy in PEGTL~\cite{PEGTL}, a widely used parser library that uses memory-mapped files. While a C program that \texttt{mmap}s an empty file executes safely on native, the Wasm version, however, crashes when running under Node.js, leaving little indication of the root cause within the original C source.
These discrepancies can even pose real-world risks: Kalign~\cite{KALIGN}, a widely used multiple sequence alignment tool in bioinformatics, quietly produces divergent outcomes for the same protein input under Wasm and native~\cite{WasmDiff}, undermining the foundation of downstream analysis.

\input{figures/DiscrepancyExample}

\begin{figure*}[t]
    \centering
    \includegraphics[width=0.95\textwidth]{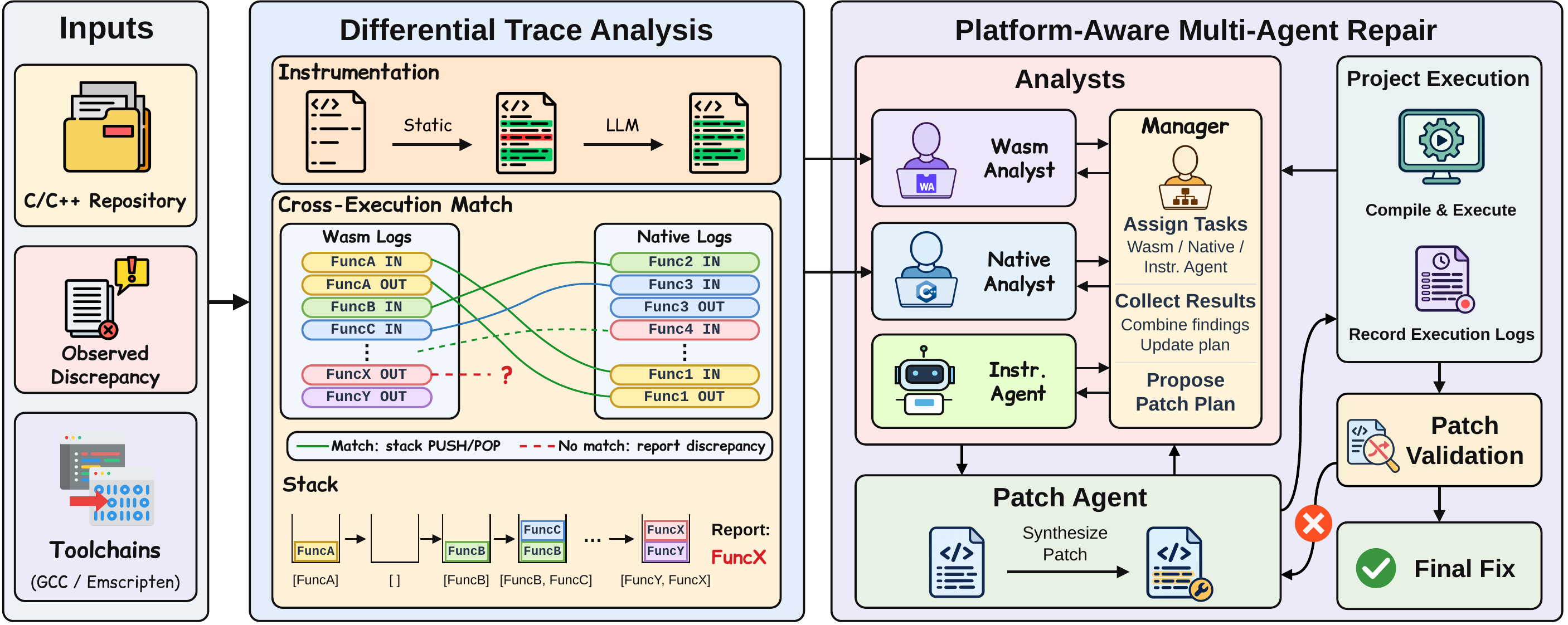}
    \caption{Architecture of \tool}
    \Description{Flowchart of divergence localization and iterative repair. A C/C++ repository, an observed discrepancy, and toolchains feed into differential trace analysis. Static and language-model-assisted instrumentation produce native and WebAssembly execution logs. Cross-execution matching connects corresponding function entry and exit events, maintains a call stack, and reports an unmatched event at FuncX. The report feeds into platform-aware repair, where a Manager assigns tasks to native and WebAssembly analysts and an Instrumentation Agent, combines their findings, and proposes a patch plan. A Patch Agent synthesizes a patch. Compilation and execution produce logs for further analysis and patch validation; failed validation returns to patch synthesis, while successful validation yields the final fix.}
    \label{fig:tool_architecture}
\end{figure*}

The high cost of manually diagnosing and repairing these discrepancies motivates the need for automated approaches. 
However, existing Large Language Model (LLM) agents struggle to address Native-Wasm functional discrepancies due to two unique challenges:
\textbf{Divergence Localization} and \textbf{Cross-Platform Overload}. 

First, although LLM-based agents have achieved strong performance on benchmarks focusing on logic mistakes such as Defects4J~\cite{Defects4J}, the complex Native-Wasm functional discrepancies often originate from environment settings, library implementations, or compiler bugs, which are hidden in the program's internal execution state~\cite{WasmChecker}. 
Our experiment showed that, in practice, without concrete guidance toward the source of divergence, existing state-of-the-art coding agents are frequently puzzled by undirected exploration inside the repositories, failing to locate the direct causes of discrepancies and fix them. 

Besides divergence localization, cross-platform overload presents another challenge. To propose a patch, the LLM agents also need to interpret execution evidence from both native and Wasm platforms. However, we observed that a single agent analyzing both platforms tended to become overwhelmed by the reasoning overhead, failing to interpret and compare the two platforms' behavior. The agent’s overconfidence in its knowledge of C/C++ led it to overlook Wasm, thus hindering its ability to align behavior across the two platforms.

To address these challenges, we propose \tool, a multi-agent repair framework that combines automated divergence localization with coordinated platform-specific analysis. The core idea is to mirror how human experts debug cross-platform discrepancies: instrument source code, run both builds, and compare execution traces to localize the divergence. Then, based on the traces, experts can investigate each platform’s behavior and integrate their findings into an explanation of the discrepancy to form a concrete patch plan.

As shown in Figure~\ref{fig:tool_architecture}, given a repository with an observed discrepancy between native and Wasm execution, \tool first performs differential trace analysis by instrumenting relevant functions and matching corresponding events in native and Wasm execution traces. This analysis localizes a suspicious divergence with its calling context, providing concrete observations and a focused starting point for investigation. 
Building on this result, \tool employs a platform-aware multi-agent repair framework to explain the discrepancy and develop a patch. 
A Manager Agent decomposes the investigation into platform-specific tasks assigned to corresponding native and Wasm analysts,
allowing each to concentrate on interpreting a single 
platform’s behavior.
When needed, an Instrumentation Agent can be invoked to collect additional observations for the analysts. The manager integrates the analysts' findings to refine its explanation of the discrepancy and formulate a patch plan. 
Guided by this plan, the Patch Agent generates a candidate patch and reviews the modified code alongside compilation and execution results from both platforms. 
If the patch fails, the resulting 
feedback guides further investigation and patch revision.

To evaluate \tool, we curate a benchmark of 114 unique, nontrivial Native-Wasm functional discrepancies.
Our benchmark targets complex discrepancies in real-world C/C++ repositories arising from library implementation differences, compiler bugs, and platform-specific behaviors in Emscripten, the most widely used Wasm toolchain~\cite{EmpiricalWasmCompilerBugs}. 
Across five backbone models, \tool achieves a fix rate of up to 84.8\% and improves the average fix rate by at least 37.2 percentage points over the evaluated open-source coding agent baselines, while maintaining comparable token usage and cost.

In summary, this work makes the following contributions:
\begin{itemize}[nosep,noitemsep,leftmargin=*]

    \item We introduce Native-Wasm functional discrepancy repair, a challenging task whose root causes are difficult to identify from source code alone, and present a benchmark comprising 114 complex discrepancy cases.

    \item We propose \tool, which combines a novel differential trace-matching algorithm for divergence localization with a multi-agent framework that utilizes platform-specific investigations to guide automated repair.

    \item We conduct an extensive evaluation across five recent LLMs, demonstrating that \tool consistently outperforms state-of-the-art open-source coding agent frameworks.

\end{itemize}

%% file: figures/DiscrepancyExample.tex
\begin{figure}[t]
\centering
\begingroup
\definecolor{bg}{RGB}{244,246,248}
\definecolor{cm}{RGB}{64,145,100}
\definecolor{shellbg}{RGB}{250,248,244}
\lstdefinestyle{figure-code}{
  style=code,
  basicstyle=\ttfamily\fontsize{7}{8}\selectfont,
  columns=fixed, basewidth=0.49em, keepspaces=true,
  lineskip=0.85pt,
  breakatwhitespace=true,
  commentstyle=\color{cm}\bfseries,
  xleftmargin=13pt, xrightmargin=5pt,
  framexleftmargin=10pt, framexrightmargin=2pt,
  aboveskip=10pt, belowskip=2pt,
}
\lstdefinestyle{figure-shell}{
  style=shell,
  columns=fixed, basewidth=0.53em, keepspaces=true,
  lineskip=0.85pt,
  breakatwhitespace=true,
  basicstyle=\ttfamily\fontsize{7}{8}\selectfont\bfseries\color{black},
  moredelim=[is][\color{shellcmd}\mdseries]{`C}{C`},
}
\captionsetup[subfigure]{skip=3pt}
\begin{subfigure}[t]{\linewidth}
\begin{lstlisting}[style=figure-code]
// main.c
int fd = open("./empty_file", O_RDWR);
struct stat st;  fstat(fd, &st);
int size = st.st_size;  // == 0
// mmap the empty file with len 0
char *p = (char*)mmap(NULL, size, PROT_READ, MAP_PRIVATE, fd, 0); 
printf("Program finished\n");
\end{lstlisting}
\caption{Minimal reproducer.}
\label{fig:mmap-code}
\end{subfigure}

\vspace{5pt}
\begin{subfigure}[t]{0.35\linewidth}
\begin{minipage}[t][42pt][t]{\linewidth}
\begin{lstlisting}[style=figure-shell]
`P$ P``Cgcc main.c -o mainC`
`P$ P``C./mainC`
Program finished
\end{lstlisting}
\end{minipage}
\caption{Native (gcc).}
\label{fig:mmap-native}
\end{subfigure}%
\hfill
\begin{subfigure}[t]{0.61\linewidth}
\begin{minipage}[t][42pt][t]{\linewidth}
\begin{lstlisting}[style=figure-shell]
`P$ P``Cemcc main.c -o main.jsC`
`P$ P``Cnode main.jsC`
`ETypeError: Cannot read properties of null (reading 'length')E`
\end{lstlisting}
\end{minipage}
\caption{Emscripten (emcc Ver 3.1.54).}
\label{fig:mmap-wasm}
\end{subfigure}

\caption{Example Native-Wasm functional discrepancy: Calling \texttt{mmap} on a zero-length file triggers a JavaScript glue code crash that never happens under native execution. (a) shows the minimal reproducer, (b) displays the native C execution, and (c) illustrates the Wasm execution via Node.js.}
\Description{Three panels show a C reproducer and its two execution outputs. The code opens an empty file, reads its size with fstat, calls mmap with that size, and then prints a completion message. Below the code, the native terminal prints Program finished, while the Emscripten terminal reports a TypeError when attempting to read the length property of null.}
\label{fig:mmap-bug}
\endgroup
\end{figure}

%% file: tex/RelatedWork.tex
\section{Related Work}
\label{sec:related_work}

\textbf{Discrepancies between Wasm and Native Executions.}
With support across all major browsers~\cite{webassembly_features}, Wasm has seen wide adoption in web applications~\cite{WasmBench, empower_webapp_with_wasm, understanding_performance_of_wasm_applications}, edge computing~\cite{wasm_edge_computing_1, wasm_edge_computing_2}, and the Internet of Things~\cite{wasm_iot_1, wasm_iot_2, wasm_iot_3}, driven by the promise of seamless code migration~\cite{WasmDiff}. However, cross-compiling C/C++ to Wasm can silently violate native semantics. While some studies~\cite{PC2W, security_risk_lack_compiler_protection} highlight security vulnerabilities using the Juliet test suite~\cite{Juliet}, others examine functional discrepancies~\cite{WasmDiff}. Notably, \citet{WasmChecker} provide a comprehensive study on real-world repositories at scale, exposing the severity of these divergences. Yet, these works stop at detection and characterization. To our knowledge, \tool is the first system to automatically fix discrepancies between Wasm and native executions.

\noindent\textbf{LLM-Based Program Repair.}
LLMs serve as effective foundations for automated program repair~\cite{LLMAPRSurvey}. Many approaches reason about bug root causes using source code logic and test signals~\cite{Agentless, TeachSelfDebug, FitRepair}, with some leveraging multi-turn conversational feedback (e.g., ChatRepair~\cite{ChatRepair}) or few-shot demonstrations (e.g., ThinkRepair~\cite{ThinkRepair}). Recent agent-based methods push automatic program repair further~\cite{AdverIntentAgent, HyperAgent}. For instance, AutoCodeRover~\cite{AutoCodeRover} integrates structure-aware code search, RepairAgent~\cite{RepairAgent} uses a state machine for tool-governed diagnosis, and MAGIS~\cite{MAGIS} coordinates role-specialized agents. 
General-purpose coding agent frameworks, including SWE-agent~\cite{SWE-Agent}, OpenCode~\cite{opencode}, and OpenHands~\cite{openhands}, provide flexible workflows for diverse software engineering tasks, including bug diagnosis and repair.
However, these works largely focus on source-level logical bugs within standard benchmarks~\cite{Defects4J, QuixBugs, SWE-bench, evalplus}. Reasoning from source alone often struggles to capture the actual program states needed for Native-Wasm discrepancy repair. Even works that combine static analysis with LLM reasoning~\cite{HybridAPR, AgenticProgramRepairfromTestFailure} or explicitly use LLM-driven instrumentation~\cite{TraceCoder, TraceRepairer, DebugRepair} to obtain internal states remain unsuitable for our setting: the former cannot expose dynamic execution divergence, and the latter lack concrete guidance for where to instrument. \tool, on the other hand, integrates a differential trace analysis that aligns native and Wasm executions and localizes the divergence, along with a multi-agent framework consisting of platform-specific agents for effective Native-Wasm discrepancy understanding and repair.

%% file: tex/Methodology/Methodology.tex
\section{Methodology}

This section details the design of \tool. We first formalize the Native-Wasm functional discrepancy repair problem (Section~\ref{ssec:problem_formulation}), and present \tool's two stages: {differential trace analysis} (Section~\ref{ssec:instrumentation_and_root_cause_localization}), which instruments source code (Section~\ref{ssec: instrumentation}) and localizes the initial divergence through cross-execution event matching (Section~\ref{ssec: cross-execution_match}); and {platform-aware multi-agent repair} (Section~\ref{ssec:divergence_guided_agentic_repair_v2}), which employs platform-specific agents for diagnosis and uses their findings to synthesize and validate patches.

\input{tex/Methodology/Formulation}

\input{tex/Methodology/DiffTraceAnalysis}

\input{tex/Methodology/AgentRepairV2}

%% file: tex/Methodology/Formulation.tex
\subsection{Problem Formulation}
\label{ssec:problem_formulation}

Given a repository $S$, a native compiler toolchain $C_n$ (e.g., GCC), and a Wasm compiler toolchain $C_w$ (e.g., Emscripten), we use native behavior as the reference.
A \textit{Native-Wasm functional discrepancy} exists when there is an input $I$ such that
\[\texttt{exec}(C_n(S), I) \not\equiv \texttt{exec}(C_w(S), I).\]
Here, $\texttt{exec}$ executes the compiled program and captures its observable behavior on a given input. $\not\equiv$ denotes a divergence in observable behaviors, such as mismatched standard outputs, differing exit codes, or one execution crashing while the other succeeds. 

A repair for the discrepancy is a source-code patch $P$ such that applying $P$ to $S$ (denoted $S \oplus P$) resolves the discrepancy while preserving the original native behavior:
$\texttt{exec}(C_n(S), I) \equiv \texttt{exec}(C_n(S \oplus P), I) \equiv \texttt{exec}(C_w(S \oplus P), I)$.
Beyond resolving the discrepancy for input $I$, the patch should preserve the original native behavior on other valid inputs: given a valid $I' \neq I$, $\texttt{exec}(C_n(S), I') \equiv \texttt{exec}(C_n(S \oplus P), I') \equiv \texttt{exec}(C_w(S \oplus P), I')$.
An example patch for the discrepancy displayed in Figure~\ref{fig:mmap-bug} is provided in Figure~\ref{fig:mmap-fix} in Appendix~\ref{app:examples}.

%% file: tex/Methodology/DiffTraceAnalysis.tex
\subsection{Differential Trace Analysis}
\label{ssec:instrumentation_and_root_cause_localization}

\subsubsection{Instrumentation}
\label{ssec: instrumentation}

As illustrated in Figure~\ref{fig:tool_architecture}, the first step of differential trace analysis is instrumentation. Leveraging call graphs and the location of the observed discrepancy (e.g., different test case outcomes), \tool automatically tracks function dependencies reachable from this location to identify all relevant types and functions in the repository and apply instrumentation, which involves the following components:

\noindent\textbf{Type Serialization.} \tool identifies all user-defined types of arguments, return values, and local variables within the relevant functions. 
It analyzes the dependencies among these types and applies a topological sort so that serialization for a member type is generated before serializing the type that contains it. Following this order, 
\tool uses \texttt{libclang} to inspect each relevant type's members in the abstract syntax tree (AST) and statically generate the corresponding serialization code.

\noindent\textbf{Function Instrumentation.} For each function, \tool inserts print statements at entry and immediately before each exit. Each message includes the function's unique ID and the values of the corresponding input or output variables. \tool generates these statements using the AST and handles user-defined types via the type-specific serialization described above.

\noindent\textbf{LLM Assistance.} Throughout the instrumentation process, \tool routinely addresses compilation errors or inconsistent test outcomes caused by the static type serialization or function instrumentation.
For example, compilation failures can occur when using the statically generated printing statement for standard library types such as \texttt{std::locale} that lack predefined output stream operators. When such errors arise, \tool isolates the problematic type or function and employs an LLM to repair the inserted instrumentation code. If the issue persists, \tool marks the affected function as \textit{concession}, where the models are allowed to omit the serialization of specific variables or, in the worst case, \tool restores the original function.

\subsubsection{Cross-Execution Match}
\label{ssec: cross-execution_match}

\tool compares the native execution trace against the Wasm trace, using the native trace as the reference. Each \textit{event} records the entry or exit of a function invocation, together with the function ID and observed variable states. Algorithm~\ref{alg:diff_trace} outlines the analysis: \tool maintains a call stack and a list of suspects associated with incomplete instrumentation, and processes native events sequentially. For each event, \tool first searches for a matching Wasm event. If a match is found, it updates the call stack and suspect list. Otherwise, \tool stops and reports the discrepancy using the accumulated diagnostic context.

\noindent \textbf{Flexible Event Match.} For each native event, \texttt{Match\-Wasm\-Event} searches the Wasm event pool for an unused event with the same function ID and entry/exit flag, together with equivalent recorded states. 
\tool allows flexible matching because behavior left unspecified by the C standard, such as argument evaluation order, can produce different function call orders.
An example is provided in Figure~\ref{fig:ub-bug} in Appendix~\ref{app:examples}.
Equivalence requires all corresponding recorded values to satisfy the comparison rules: ordinary values are compared directly, whereas platform-dependent values, such as memory addresses and file descriptors, are normalized before comparison, to avoid treating differences in low-level representations as execution discrepancies. Once a match is found, the Wasm event is marked as used, ensuring that repeated native invocations cannot reuse the same Wasm event. If no unused candidate satisfies these conditions, the native event is considered unmatched. Appendix~\ref{appendix:supp_of_diff_trace_analysis} provides the detailed matching algorithm.

\noindent \textbf{Stack-Based Localization.} \tool reconstructs a call stack while processing matched native events: a matched entry pushes the corresponding event onto \texttt{Stack}, and a matched exit pops the top entry. At the first unmatched native event, Algorithm~\ref{alg:diff_trace} invokes \texttt{Report(Stack, Suspects, evt)} and terminates without updating the stack for that event. 
For an entry mismatch, our analysis points to the caller, which supplied the function's arguments. The caller is at the top of the stack because the unmatched entry has not been pushed. For an exit mismatch, the analysis points to the exiting function, which remains at the top. Therefore, the function represented by the top stack entry is always reported as the ``initial divergence location''.

\begin{algorithm}[t]
        \setstretch{0.9}%
            \newcommand\mycommfont[1]{\textcolor{deepgreen}{#1}}
            \SetCommentSty{mycommfont}
            \DontPrintSemicolon
            \caption{Cross-Execution Match}
            \label{alg:diff_trace}
            \small 
            
            \KwIn{\texttt{NativeEvents, WasmEvents}: Sequential streams of instrumented entry/exit events from native/Wasm execution}
            \KwOut{Root-cause discrepancy report}
            \texttt{Stack} $\gets []$\;
            \texttt{Suspects} $\gets []$\;
            \ForAll{\texttt{evt} $\in$ \texttt{NativeEvents}}{
                \texttt{is\_matched} $\gets$ \texttt{MatchWasmEvent(evt, WasmEvents)}\;
                \eIf{\texttt{is\_matched}}{
                    \uIf{\texttt{IsEntry(evt)}}{
                        \texttt{Stack.Push(evt)}\;
                    }
                    \ElseIf{\texttt{IsExit(evt)}}{
                        \texttt{Stack.Pop()}\;
                    }
                    \texttt{UpdateSus(evt, Stack, Suspects)}\;
                }{
                    \tcp{Entry or exit mismatch}
                    \Return \texttt{Report(Stack, Suspects, evt)}\;
                }
            }
\end{algorithm}

\noindent \textbf{Tracking Incomplete Instrumentation.} Instrumentation \textit{concessions} (defined in Section~\ref{ssec: instrumentation}) can leave execution states unobserved, limiting localization based on the stack alone. 
We therefore design \texttt{UpdateSus} (Algorithm~\ref{alg:update_sus}) to track function invocations affected by incomplete instrumentation as suspects.
The tracker is invoked after the call stack is updated for each matched event.
If a matched entry event belongs to an invocation with concessions, the tracker adds that invocation to \texttt{Suspects} and records its caller as its \textit{witness} in the persistent map \texttt{WitnessMap}.
On a matched exit, the tracker retrieves all suspects whose current witness is the exiting invocation. If the exiting invocation is flagged as concession, the suspects are retained, and their witness becomes the exiting invocation's caller. Otherwise, the matched exit of the witness releases the associated suspects, removing both the suspects' entries in \texttt{Suspects} and their mappings in \texttt{WitnessMap}. In other words, a suspect's own matched exit does not clear the suspect, since its unrecorded states may still differ. Instead, the clearance depends on a matched exit of its assigned witness without concessions. 
If an invocation has no caller when a witness is assigned or updated, the associated suspects will remain in \texttt{Suspects} and never be removed.

\noindent \textbf{Diagnostic Report.} When a mismatch occurs, \texttt{Report} terminates the analysis and collects the unmatched event, the current stack, unresolved suspects, available native and Wasm I/O states, and relevant variable-type definitions. Together, the stack and unresolved suspects provide repair agents with both the immediate mismatch context and earlier invocations that remain potentially relevant due to incomplete instrumentation.

%% file: tex/Methodology/AgentRepairV2.tex
\subsection{Platform-Aware Multi-Agent Repair}
\label{ssec:divergence_guided_agentic_repair_v2}

\begin{algorithm}[t]
    \setstretch{0.9}%
    \providecommand\mycommfont[1]{\textcolor{deepgreen}{#1}}
    \SetCommentSty{mycommfont}
    \SetKwInput{KwState}{Global}
    \DontPrintSemicolon
    \caption{Suspect Tracker (\texttt{UpdateSus})}
    \label{alg:update_sus}
    \small

    \KwIn{\texttt{evt}: Current matched entry/exit event \\
          \texttt{Stack}: Updated call stack \\
          \texttt{Suspects}: Current suspect invocations}
    \KwOut{Updated suspects and witness map}
    \KwState{\texttt{WitnessMap}: Persistent suspect-to-witness map}

    \texttt{inv} $\gets$ \texttt{Invocation(evt)} \tcp{Get event's associated invocation}
    \If{\texttt{IsEntry(evt)} $\land$ \texttt{HasConcession(inv)}}{
        \tcp{Register suspect and caller witness}
        \texttt{Suspects.Append(inv)}\;
        \texttt{WitnessMap[inv]} $\gets$ \texttt{Caller(inv, Stack)}\;
    }
    \If{\texttt{IsExit(evt)}}{
        \texttt{asso\_sus} $\gets$ \texttt{WitnessMap.KeysFor(inv)}\;
        \eIf{\texttt{HasConcession(inv)}}{
            \tcp{Transfer to caller; None retains suspects}
            \ForAll{\texttt{sus} $\in$ \texttt{asso\_sus}}{
                \texttt{WitnessMap[sus]} $\gets$ \texttt{Caller(inv, Stack)}\;
            }
        }{
            \tcp{Clean witness releases associated suspects}
            \texttt{Suspects.RemoveAll(asso\_sus)}\;
            \texttt{WitnessMap.DeleteKeys(asso\_sus)}\;
        }
    }
\end{algorithm}

With the divergence localized by differential trace analysis, \tool uses a multi-agent framework to understand the discrepancy and synthesize patches. A Manager Agent coordinates native and Wasm analysts and on-demand instrumentation to develop a patch plan that guides subsequent patch generation and validation.

\noindent\textbf{Manager Agent.}
The Manager Agent coordinates the investigation, integrates findings, and maintains the patch plan. Starting from trace analysis and discrepancy metadata, it forms an initial hypothesis about why the discrepancy occurs. To assess and refine this hypothesis, the manager assigns questions to the native or Wasm analyst, asking it to interpret the relevant platform behavior based on observations from trace analysis or the instrumentation agent. The analysts' returned findings are checked against the available evidence to determine whether the hypothesis explains the observed discrepancy and to identify any remaining unresolved questions. 
The manager then decides whether to
ask an analyst a follow-up question or request the Instrumentation Agent to collect intermediate variable states within relevant functions. 
Once the evidence is sufficient, the manager formulates a patch plan that includes a description of the suspected root cause supported by evidence and the proposed code modifications. The plan explains why these changes should resolve the discrepancy. The manager then passes the plan to the Patch Agent.
When new observations or unsuccessful patch attempts indicate that the plan needs revision, the manager repeats the relevant investigation steps and updates its hypothesis and patch plan accordingly.

\noindent\textbf{Platform-Specific Analysts.}
The Manager Agent delegates detailed investigation of platform behavior to separate native and Wasm analysts. Each analyst accesses the relevant source code and the corresponding platform’s execution records from differential trace analysis or the instrumentation agent to investigate how the code interacts with the platform’s compiler, libraries, and runtime to produce the observed behavior. Throughout this investigation, the analyst assesses whether execution traces and source code support the manager's hypothesis, and suggests alternatives when needed. Each analyst returns a structured report explaining the observed platform behavior with the supporting evidence. The report clarifies the discrepancy and identifies unresolved assumptions, allowing the manager to compare and integrate the platform-specific explanations, direct further evidence collection, and refine the patch plan.

\noindent\textbf{Instrumentation Agent.}
Unlike the initial function-level instrumentation used in differential trace analysis, the Instrumentation Agent performs targeted, on-demand instrumentation during repair to capture additional intermediate execution states within a function, such as local variable values, on both native and Wasm platforms.
The agent inserts print statements to capture these states, labeling each state with the location in the original function. The framework then builds and executes the instrumented program on both platforms, collects the recorded states, and restores the source after each instrumentation attempt. The captured states and their source locations are made available to the manager and the corresponding platform analysts, helping them to examine how values are produced and transformed within the function and assess explanations of the observed discrepancy.

\noindent\textbf{Patch Generation and Validation.}
The Manager Agent passes the established patch plan to the Patch Agent, which generates a candidate patch that can span multiple source locations or files. The framework applies all edits together, rebuilds the program, and executes it on both native and Wasm platforms. The Patch Agent reviews the modified code alongside the compilation and execution results to assess whether the patch addresses the underlying discrepancy and preserves the intended functionality. The review also rejects patches that circumvent validation, for example, by skipping test cases. 
Unsuccessful or rejected patches are automatically reverted.
The Patch Agent can then revise the patch under the existing plan or return control to the Manager Agent when the feedback indicates a need for further investigation or plan revision.

%% file: tex/Evaluation/Evaluation.tex
\section{Evaluation}

In this section, we evaluate \tool's performance in repairing Native-Wasm functional discrepancies.
After presenting the experimental setup (Section~\ref{ssec:eval_setup}), we compare \tool with three open-source coding agents in terms of repair effectiveness and cost efficiency (Sections~\ref{ssec:effectiveness_of_wasmmend} and~\ref{ssec:efficiency_of_tool}). We then evaluate the localization quality of differential trace analysis (Section~\ref{ssec: trace_analysis_quality}) and examine the contributions of the repair framework and trace analysis (Section~\ref{ssec:contrib_of_repair_framework_and_trace_analysis}). Finally, we analyze the cases that \tool fails to repair (Section~\ref{ssec:failure_analysis}).

\input{tex/Evaluation/Exp_setup}

\input{tex/Evaluation/Effectiveness}

\input{tex/Evaluation/Efficiency}

\input{tex/Evaluation/TraceAnalysisQuality}

\input{tex/Evaluation/DiffTraceQuality}

\input{tex/Evaluation/FailureAnalysis}

%% file: tex/Evaluation/Exp_setup.tex
\subsection{Evaluation Setup}
\label{ssec:eval_setup}

\noindent\textbf{Dataset.}
\tool targets complex discrepancies in C/C++ programs, the predominant source languages for Wasm~\cite{WasmBench}. In particular, following WasmChecker's classification~\cite{WasmChecker}, we focus on \textit{true test discrepancies}, i.e., cases that cannot be resolved by adjusting compilation flags and therefore require source-code adaptation. 
We start with the discrepancy cases reported by WasmChecker and follow its strategy to collect new candidates. We crawl additional GitHub repositories and execute their built-in unit tests using both native and Wasm builds to identify Wasm-specific failures as discrepancy candidates. 
We then filter both the WasmChecker discrepancies and the newly discovered candidates, retaining only \textit{true test discrepancies}. When potential duplicates arise within the same repository, we check whether they share a root cause and retain only one case if they do.
This yields 114 unique Native-Wasm functional discrepancies (23 from WasmChecker and 91 newly discovered), each surfaced by unit tests in real-world C/C++ repositories. The statistics of the dataset are available in Appendix~\ref{appendix:dataset_categorization}.

\noindent\textbf{Baseline Frameworks.} To evaluate the effectiveness and efficiency of \tool, we compare it with three open-source coding agent frameworks, OpenCode~\cite{opencode}, OpenHands~\cite{openhands}, and SWE-agent~\cite{SWE-Agent}, on the Native-Wasm functional discrepancy repair task. Each framework is given the unit test on which the native and Wasm execution results diverge, along with access to every file available to \tool. We also grant each baseline 50\% more iterations than \tool, giving them more opportunities to produce a fix.

\noindent\textbf{Backbone Models.} We evaluate \tool and the baseline agent frameworks on five models released after May 2026: \DS (Sept. 2026)~\cite{deepseekv4.1}, \Mimo (Sept. 2026)~\cite{mimo-v2.6}, \Minimax (June 2026)~\cite{minimax-m3}, \Glimmer (Aug. 2026)~\cite{muse-glimmer}, and \Qwen (Aug. 2026)~\cite{qwen3.8flash}.

\noindent\textbf{Repair Setup and Success Criterion.}
For differential trace analysis, we reuse the results obtained using DeepSeek-V4.1-Flash across all runs, regardless of the backbone model used for repair. This decision is based on two reasons: 
1) the five evaluated models achieve similar trace analysis quality on a subset of 40 cases\footnote{The statistics are provided in Table~\ref{tab:llm-differential-trace-analysis}.}; 
2) reusing the same trace analysis results helps to isolate the impact of the backbone model on discrepancy repair.
In our experiments, we allow \tool to execute at most 50 iterations, where every LLM call, including those made by the platform-specific analysts and the instrumentation agent, counts as one iteration. As discussed above, the open-source coding agents are allowed up to 75 iterations. During repair, the agents can access only the unit test that exposes the discrepancy, and all other tests in the same suite are withheld. This restriction aims to prevent hardcoded patches and test skipping. The withheld test cases are used only to evaluate whether the proposed patch fixes the discrepancy.
A discrepancy is considered fixed if the patched program passes all unit tests in the suite under both native and Wasm execution, which reflects the repair objective in Section~\ref{ssec:problem_formulation}.
We report fix rates as averages over three runs.

%% file: tex/Evaluation/Effectiveness.tex
\subsection{Effectiveness of \tool}
\label{ssec:effectiveness_of_wasmmend}

\input{tables/Effectiveness}

Table~\ref{tab:effectiveness} compares \tool with the three baseline frameworks across five backbone models. \tool achieves an overall fix rate of 69.5\%, more than twice the highest baseline average, demonstrating its strong capability for patching Native-Wasm discrepancies.

We first check whether the cases fixed by \tool form a superset of those fixed by the baselines. For most backbone models, \tool fixes over 98\% of the cases that each baseline fixes. The only exception is \Qwen, where \tool fixes 97\% of the cases fixed by OpenCode, 85\% of those fixed by OpenHands, and 88\% of those fixed by SWE-agent. 
To better understand why \tool resolves discrepancies that the baselines fail to repair, we further examine how its design contributes to these successful repairs.

Open-source coding agents often spend many iterations investigating the root cause of a discrepancy. Without a systematic approach like differential trace analysis, they fall back on general-purpose exploration: reading the test, grepping for symbols, and inspecting candidate files one by one, spending many iterations before touching the file that actually needs to change. \tool instead begins with trace analysis to narrow the search to a small set of candidate files. In case studies, \tool typically begins analyzing and patching the root-cause files within 20 iterations, whereas the open-source agents do so only after 37 iterations on average.  Faster localization also matters beyond the final fix rate: the iterations the baselines spend searching consume context budget and LLM calls, and in 59\% (OpenCode 66\%, OpenHands 31\%, SWE-agent 80\%) of failed baseline runs the agent exhausts its iteration limit without ever editing the root-cause file. These patterns also help explain the particularly low baseline fix rates with \Mimo and \Qwen: both models frequently get stuck in exploration, leaving few or no iterations for repository edits.

In addition, by using a dedicated agent that focuses on Wasm, \tool can more efficiently identify files that affect Wasm behavior. Take \DS as an example: over 99\% of the files that \tool analyzes and patches contain code exercised during Wasm execution, whereas 33-38\% of the files analyzed and patched by the baselines are irrelevant to Wasm execution. A closer look reveals that most of these irrelevant accesses stem from the agents' overconfidence in their knowledge of C/C++: the baseline agents are drawn to native libraries that have no bearing on the Wasm build, which misleads them and wastes a substantial number of LLM calls.

Beyond these two design choices, we observe that \tool's patch validation workflow helps prevent unsuccessful changes from accumulating by reverting failed patches while retaining their execution logs for the next attempt.
By contrast, across all baselines, unsuccessful patches sometimes accumulate, complicating later repairs and causing both native and Wasm tests to continue failing even after the original cause is resolved. 
We hypothesize that the complexity of diagnosing and repairing Native-Wasm discrepancies can lead baseline agents to overlook the need to revert unsuccessful changes between attempts.

To complement the test-based evaluation, we manually review the quality of 140 test-passing \tool patches, as detailed in Appendix~\ref{appendix:manual_fix_analysis}.

%% file: tables/Effectiveness.tex
\begin{table}[t]
\centering
\small
\caption{Fix rate (\%) of three baseline frameworks and \tool across five backbone models. Underline marks the best baseline; values in parentheses give the absolute improvement of \tool over the best baseline. \tool achieves the highest fix rate on all five models.}
\label{tab:effectiveness}
\setlength{\tabcolsep}{3pt}
\begin{tabular}{lccc|c}
\toprule
Model & OpenCode & OpenHands & SWE-agent & \textbf{WasmMend} \\
\midrule
DeepSeek-Flash & 58.8 & \underline{61.4} & 43.9 & \textbf{84.8} (+23.4) \\
MiMo-Flash & 0.0 & 1.8 & \underline{3.5} & \textbf{63.4} (+59.9) \\
MiniMax-M3 & 30.7 & \underline{31.6} & 28.1 & \textbf{68.1} (+36.5) \\
Muse-Glimmer & \underline{55.3} & 41.2 & 19.3 & \textbf{72.0} (+16.7) \\
Qwen-Flash & 16.7 & \underline{18.4} & 16.7 & \textbf{59.4} (+41.0) \\
\midrule
Average & \underline{32.3} & 30.9 & 22.3 & \textbf{69.5} (+37.2) \\
\bottomrule
\end{tabular}
\end{table}

\begin{table*}[t]
\centering
\small
\caption{Average token usage and API cost (USD) per successful fix.
New and Cached denote new and cached input tokens.
Out denotes output tokens. Cost for \tool includes the cost of Differential Trace Analysis (\$0.0018 per case). \tool incurs comparable cost on the fixed cases while achieving much higher fix rates. The average token usage and API cost on all cases are reported in Table~\ref{tab:tool_efficiency_all_cases} in Appendix~\ref{appendix:full_efficiency}.}
\label{tab:tool_efficiency}
\setlength{\tabcolsep}{4pt}
\renewcommand{\arraystretch}{1.20}

\begin{tabular*}{0.98\textwidth}
    {@{}l@{\hspace{28pt}}
     *{2}{c@{\extracolsep{\fill}\hspace{4pt}}r@{\extracolsep{0pt}\hspace{18pt}}}
     c@{\extracolsep{\fill}\hspace{4pt}}r@{\extracolsep{0pt}\hspace{9pt}}
     |@{\hspace{9pt}}c@{\extracolsep{\fill}\hspace{4pt}}r@{\extracolsep{0pt}}}
\toprule
\multirow{2}{*}{Model}
& \multicolumn{2}{@{}c@{\hspace{18pt}}}{OpenCode}
& \multicolumn{2}{@{}c@{\hspace{18pt}}}{OpenHands}
& \multicolumn{2}{@{}c@{\hspace{\dimexpr18pt+\arrayrulewidth\relax}}}{SWE-agent}
& \multicolumn{2}{@{}c@{}}{\textbf{WasmMend}} \\
\cmidrule(r{18pt}){2-3}\cmidrule(r{18pt}){4-5}
\cmidrule(r{\dimexpr18pt+\arrayrulewidth\relax}){6-7}\cmidrule{8-9}
& New/Cached/Out & Cost
& New/Cached/Out & Cost
& New/Cached/Out & Cost
& New/Cached/Out & Cost \\
\midrule

DeepSeek-Flash
& 34K/1,341K/12K & 0.033
& 42K/1,160K/16K & 0.039
& 28K/1,067K/13K & 0.030
& 78K/720K/19K & 0.052 \\

MiMo-Flash
& N/A & N/A
& 78K/750K/6K & 0.015
& 43K/1,041K/10K & 0.012
& 127K/1,253K/12K & 0.026 \\

MiniMax-M3
& 61K/1,294K/9K & 0.107
& 91K/1,459K/26K & 0.147
& 63K/1,281K/11K & 0.108
& 165K/883K/11K & 0.118 \\

Muse-Glimmer
& 68K/1,313K/9K & 0.090
& 79K/1,077K/14K & 0.091
& 42K/464K/10K & 0.049
& 135K/591K/11K & 0.089 \\

Qwen-Flash
& 48K/1,273K/9K & 0.032
& 65K/1,069K/10K & 0.032
& 71K/1,655K/11K & 0.042
& 91K/639K/8K & 0.029 \\

\midrule
Average
& 53K/1,305K/10K & 0.065
& 71K/1,103K/14K & 0.065
& 49K/1,101K/11K & 0.048
& 119K/817K/12K & 0.063 \\
\bottomrule
\end{tabular*}
\end{table*}

%% file: tex/Evaluation/Efficiency.tex
\subsection{Efficiency of \tool}
\label{ssec:efficiency_of_tool}

\begin{figure*}[t]    \centering
    \includegraphics[width=\linewidth]{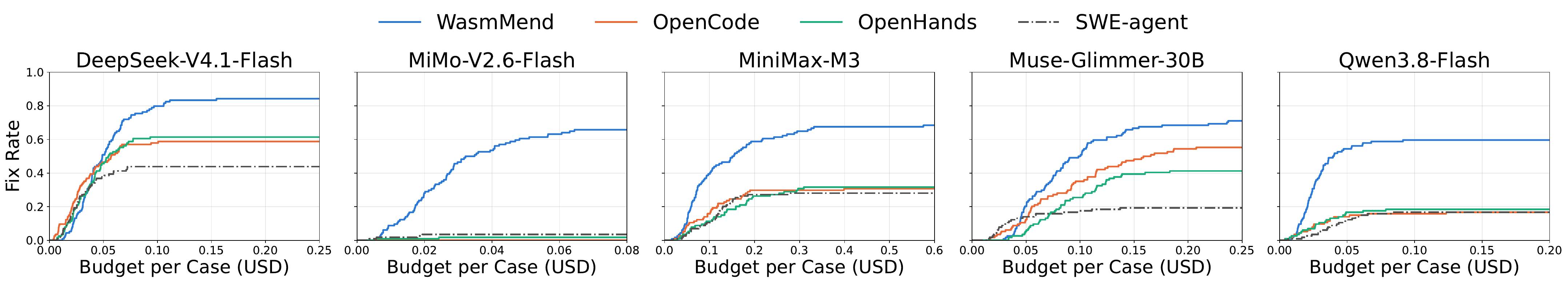}
    \caption{Fix rate with respect to inference budget (USD). \tool (\textcolor{blue}{blue}) achieves higher fix rates than the baselines at comparable budgets in most scenarios and attains the highest fix rate for every evaluated LLM.}
    \Description{Five line charts compare repair success as the budget per case increases. From left to right, the panels show DeepSeek, MiMo, MiniMax, Muse-Glimmer, and Qwen. Horizontal axes show budget per case in US dollars with different ranges across panels; the shared vertical scale shows fix rate from zero to one. Solid blue, orange, and green lines represent WasmMend, OpenCode, and OpenHands, respectively; a gray dash-dot line represents SWE-agent. The curves rise in steps and then plateau. The separation is especially pronounced for MiMo and Qwen: all baseline curves remain below a fix rate of 0.2, while WasmMend exceeds 0.5.}
    \label{fig:efficiency_figure}
\end{figure*}

Table~\ref{tab:tool_efficiency} shows that \tool incurs a comparable cost on successfully fixed cases. Including Differential Trace Analysis, its average cost is \$0.063: slightly lower than OpenCode and OpenHands, but higher than SWE-agent. 
Figure~\ref{fig:efficiency_figure} shows that although some baselines are competitive at low budgets, \tool continues to improve as the budget increases, even after baseline fix rates have plateaued. This sustained improvement translates into higher fix rates over most matched budgets.
For example, with \Qwen at a budget of \$0.10 per case, \tool achieves a fix rate of approximately 60\%, while all baselines remain below 20\%.

\tool exhibits a distinct token usage pattern: in general, it consumes more new input tokens, substantially fewer cached input tokens, and a similar number of output tokens, resulting in lower average total input-token usage. Execution analysis reveals the underlying cause of this pattern: \tool introduces larger context updates into its early conversation by ingesting both source files and reports from the platform-specific analysts. These large contexts account for the majority of \tool's new input tokens. This upfront investment in understanding the discrepancy pays off soon: \tool needs fewer patch attempts and shorter repair trajectories than the baselines. As a result, the accumulated conversation history is resubmitted fewer times, reducing the total number of cached input tokens. In contrast, the baselines take more steps with smaller incremental updates, repeatedly reprocessing their growing histories. 
By dedicating more tokens early on to understanding the repository context, \tool resolves issues in far fewer iterations. 

Cache pricing also influences how token savings translate into API cost. The baseline frameworks benefit more from cache discounts because cached tokens constitute a larger share of their input usage. For instance, cached input with \DS costs only 2\% of the price of new input, compared with approximately 11\% with \Qwen. This difference in cache discounts explains why \tool costs more than the baselines with \DS despite processing fewer total input tokens, yet costs less than all three baselines with \Qwen. 
As a result, \tool's token usage pattern can be more cost-effective when cached input is less discounted.

%% file: tex/Evaluation/TraceAnalysisQuality.tex
\subsection{Trace Analysis Quality}
\label{ssec: trace_analysis_quality}

\begin{table}[t]
\small
  \centering
\caption{Performance of different LLMs in differential trace analysis on a sample of 40 cases. \textit{Precise} indicates that the root cause is identified directly; \textit{Report as Suspect} indicates that the root cause is included in a suspect list of length $\leq 10$; and \textit{Missing} denotes all remaining cases.}
\label{tab:llm-differential-trace-analysis}
  \setlength{\tabcolsep}{8pt}

  \begin{tabular}{lccc}
    \toprule
    Model & Precise & Report as Suspect & Missing \\
    \midrule
    Static-Only & 24 & 9 & 7 \\
    \midrule
    DeepSeek-Flash    & 32 & 7 & 1 \\
    MiMo-Flash    & 31 & 7 & 2 \\
    MiniMax-M3        & 32 & 6 & 2 \\
    Muse-Glimmer & 32 & 6 & 2 \\
    Qwen-Flash    & 31 & 6 & 3 \\

    \bottomrule
  \end{tabular}
\end{table}

We also evaluate the quality of differential trace analysis across different backbone models on a randomly selected subset of 40 discrepancies. 
Three computer science graduate students with years of experience in both C/C++ and Wasm manually inspected each case and annotated the root cause.
With these annotations as references, we categorize trace analysis outcomes into three levels: \textit{Precise}, where the analysis directly pinpoints the root cause; \textit{Report as Suspect}, where the root cause is included within the list of suspects, and the number of suspects is small ($\leq 10$); and \textit{Missing}, where the analysis either fails to capture the root cause or the root cause is among the suspects, but the list is excessively long.

Table~\ref{tab:llm-differential-trace-analysis} shows the trace analysis results obtained without LLM assistance (static-only) and with assistance from each of the five backbone LLMs. Notably, static-only differential trace analysis achieves strong baseline performance, precisely identifying 60\% of the root causes and missing only 7 cases. 

Incorporating LLMs into differential trace analysis primarily improves instrumentation flexibility and coverage. Specifically, LLMs excel at serializing complex types and dynamic variables (e.g., \texttt{void*} pointers). The additional flexibility enables our pipeline to instrument a broader range of types and functions, reducing instrumentation \textit{concessions}, and consequently shrinking the suspect list to identify the exact root cause. Notably, all additional Precise cases gained by using LLMs come from the Report as Suspect category in the static-only baseline, while the number of Missing cases drops as the suspect lists become shorter.

Furthermore, performance varies minimally across the five evaluated LLMs. Small differences typically stem from isolated instances where one model fails to fully instrument a specific function that another model handles successfully. 
To isolate the impact of the backbone LLMs on the subsequent discrepancy repair stage, we consistently use the trace analysis obtained with \DS across all repair experiments.

%% file: tex/Evaluation/DiffTraceQuality.tex
\subsection{Contributions of the Repair Framework and Trace Analysis}
\label{ssec:contrib_of_repair_framework_and_trace_analysis}

\begin{table}[t]
  \centering
  \caption{Fix rates (\%) and costs per successful fix for OpenCode,
    WasmMend w/o Trace Analysis, and WasmMend (default).}
  \label{tab:wasmmend-comparison}
  \small
  \setlength{\tabcolsep}{2.5pt}
  \begin{tabular}{@{}l rr rr rr@{}}
    \toprule
    \multirow{2}{*}{Model}
      & \multicolumn{2}{c}{OpenCode}
      & \multicolumn{2}{c}{\shortstack{WasmMend w/o \\ Trace Analysis}}
      & \multicolumn{2}{c}{\shortstack{WasmMend\\(default)}} \\
    \cmidrule(lr){2-3}\cmidrule(lr){4-5}\cmidrule(l){6-7}
      & \shortstack{Fix Rate} & \shortstack{Cost(\$)}
      & \shortstack{Fix Rate} & \shortstack{Cost(\$)}
      & \shortstack{Fix Rate} & \shortstack{Cost(\$)} \\
    \midrule
    DeepSeek-Flash   & 58.8 & 0.033 & 70.7 & 0.042 & 84.8 & 0.052 \\
    MiMo-Flash  &  0.0 & N/A   & 39.5 & 0.023 & 63.4 & 0.026 \\
    MiniMax-M3      & 30.7 & 0.107 & 60.2 & 0.102 & 68.1 & 0.118 \\
    Muse-Glimmer & 55.3 & 0.090 & 61.4 & 0.082 & 72.0 & 0.089 \\
    Qwen-Flash    & 16.7 & 0.032 & 50.9 & 0.027 & 59.4 & 0.029 \\
    \midrule
    Average & 32.3 & 0.065 & 56.5 & 0.055 & 69.5 & 0.063 \\
    \bottomrule
  \end{tabular}
  
\end{table}

To assess the contributions of differential trace analysis and the repair framework, we compare OpenCode, \tool without differential trace analysis, and the complete \tool system, as shown in Table~\ref{tab:wasmmend-comparison}. We select OpenCode because it achieves the highest average fix rate across backbone models among the evaluated open-source baselines in Table~\ref{tab:effectiveness}. 
Across all five backbone models, \tool w/o Trace Analysis achieves substantially higher fix rates than OpenCode, and adding differential trace analysis yields further gains in the complete \tool system\footnote{Appendix~\ref{appendix:addtional_ablation_studies} reports a complementary comparison with OpenCode augmented with differential trace analysis.}. The complete system incurs a slightly higher average cost on successfully repaired cases than \tool w/o Trace Analysis, but its cost is comparable to that of OpenCode. To better understand the improvements in fix rate, we further analyze the repair process for the 40 cases with root-cause annotations described in Section~\ref{ssec: trace_analysis_quality}.

We investigate how our trace analysis improves localization by measuring the average number of iterations used to access the annotated root cause. The complete \tool system first modifies the annotated root-cause function after an average of 19.0 iterations, compared with 25.6 for \tool w/o Trace Analysis. It also uses approximately 34\% fewer iterations than \tool w/o Trace Analysis to first read the annotated root-cause function. These observations support that trace analysis helps direct agents toward relevant code earlier in the repair process, and on-demand instrumentation alone does not fully replace the localization guidance from trace analysis. 

We further analyze how \tool's coordinated platform-specific analysis helps agents interpret the discrepancy and develop repairs. Among runs that produce a successful repair, \tool w/o Trace Analysis completes the repair in an average of 20.3 additional iterations after the first explicit read of the annotated root-cause function, compared with 34.2 for OpenCode. It also averages 33\% fewer subsequent patch attempts than OpenCode. These reductions are consistent with a clearer understanding of the root cause, guiding more targeted repair decisions.

%% file: tex/Evaluation/FailureAnalysis.tex
\subsection{Failure Analysis}
\label{ssec:failure_analysis}

Despite \tool's higher discrepancy fix rates, some cases remain unresolved. Our analysis identifies two main challenges in unsuccessful repair runs: strategy convergence and platform complexity.
Strategy convergence describes difficulties in converting investigation and multiple hypotheses into a validated implementation. In some failed runs, competing hypotheses remained unresolved, and the agent either committed to a repair without sufficient supporting evidence or continued investigating until exhausting its budget. 
Moreover, repetitive investigative actions consumed the available budget without providing additional valuable evidence. Of the failed cases, 14.9\% did not attempt a patch, and 55.2\% exhibited repetitive investigative behavior.
Platform complexity describes cases where differences in library or runtime behavior make it difficult to reproduce native behavior in Wasm. Resolving them requires broader implementation changes and may even go beyond local source edits. 
For example, a discrepancy within the \texttt{fmt} project~\cite{fmt} stems from a complex bug in Emscripten's internal implementations of \texttt{dup} and \texttt{pipe}~\cite{emscripten_issue_22030}. Even when the underlying cause is identified, the agent can still struggle to produce an effective patch. 
Addressing these remaining failure patterns is our immediate next step.

%% file: tex/Appendix/ThreatsToValidity.tex
\section{Threats to Validity}
\label{sec:threats_to_validity}

\textbf{External Validity.} Our evaluation focuses on C/C++ programs compiled to Wasm using Emscripten. The benchmark combines WasmChecker cases with newly discovered discrepancies exposed by built-in unit tests in real-world repositories. Our findings are most directly applicable to these settings. Within this benchmark, our localization and repair process analyses use manually annotated root-cause locations for 40 cases. This subset may not fully represent the complete benchmark. We consider LLM-driven automated detection of additional discrepancies and evaluation across other source languages and toolchains for future work. 

\noindent\textbf{Internal Validity.} LLM-based repair outcomes can vary across runs. To account for this variability, we set the model temperature to zero and report \tool's fix rates as averages over three runs, although individual runs may still differ. Potential data contamination is another consideration, as the evaluated repositories are publicly available and may have appeared in LLM training data. However, for our newly proposed task of Native-Wasm discrepancy repair, prior familiarity with the underlying repositories does not necessarily provide a solution. Differential trace analysis may also mislocalize the divergence. For example, the flexible match may misidentify the first divergent event. Nevertheless, for the 40 annotated cases in Section~\ref{ssec: trace_analysis_quality}, the analysis only misses the root cause for one case, suggesting such mislocalization is rare.

\noindent\textbf{Construct Validity.} 
Source-level adaptations can fulfill our repair objective without compiler or runtime changes. We leave automated patching in compiler source code to future work.
We assess repair success using a test-based criterion, including regression testing under both native and Wasm execution. The reliability of this assessment depends on the coverage and assertions of the available test suites. 

%% file: tex/Conclusion.tex
\section{Conclusion}

In this paper, we propose and define the Native-Wasm functional discrepancy repair task, a challenging problem for existing LLM-agent frameworks. To address the challenge, we introduce \tool, which combines differential trace analysis to localize execution divergence and its calling context, with a platform-aware multi-agent framework that investigates platform-specific behavior and guides patch generation and validation. Our evaluation demonstrates that \tool consistently outperforms baseline coding agents while maintaining comparable repair costs. The results highlight the value of grounding repair decisions in execution evidence and coordinating platform-specific reasoning to resolve discrepancies whose causes are difficult to identify from source code alone.

%% file: tex/Appendix/Appendix.tex
\appendix

\input{tex/Appendix/Reproducibility}

\input{tex/Appendix/PatchExample}

\input{tex/Appendix/AddtionalExp}

\input{tex/Appendix/DatasetCategorization}
\newpage

\input{tex/Appendix/AdditionalInfoforMethodology}

%% file: tex/Appendix/Reproducibility.tex
\clearpage
\section{Reproducibility}
\label{appendix:reproducibility}

We will release the codebase at \url{https://github.com/wasmmend/wasmmend_arxiv}.
In the experiments, following WasmChecker's setting, we use GCC 11.4.0 and Emscripten 3.1.54. The implementation requires API keys for \DS, \Mimo, and \Qwen, as well as a TogetherAI API key for \Minimax and \Glimmer. The authors spent approximately \$1328.96\footnote{During our experiment, the DeepSeek API offered a 50\% discount during off-peak hours. Actual cost may differ.} on development and evaluation.
All budget calculations in Section~\ref{ssec:efficiency_of_tool} are based on the pricing from 
the API providers’ pricing documentation:

\urldef{\qwenurl}\url{https://modelstudio.console.alibabacloud.com/ap-southeast-1/model/market/detail/qwen3.8-flash?serviceSite=international&ref=list}
\urldef{\minimaxurl}\url{https://www.together.ai/models/minimax-m3}
\urldef{\mimourl}\url{https://mimo.mi.com/models/en-US/mimo-v2.6-flash}
\urldef{\glimmerurl}\url{https://www.together.ai/models/muse-glimmer}

\begin{itemize}
    \item \DS: \$0.3/1M New Input Tokens, \$0.006/1M Cached Input Tokens, \$1.2/1M Output Tokens. \footnote{\url{https://api-docs.deepseek.com/quick_start/pricing}}

    \item \Mimo: \$0.14/1M New Input Tokens, \$0.0028/1M Cached Input Tokens, \$0.28/1M Output Tokens. \footnote{\mimourl}

    \item \Minimax: \$0.3/1M New Input Tokens, \$0.06/1M Cached Input Tokens, \$1.2/1M Output Tokens. \footnote{\minimaxurl}

    \item \Glimmer: \$0.35/1M New Input Tokens, \$0.04/1M Cached Input Tokens, \$1.5/1M Output Tokens. \footnote{\glimmerurl}

    \item \Qwen: \$0.15/1M New Input Tokens, \$0.016/1M Cached Input Tokens, \$0.47/1M Output Tokens. \footnote{\qwenurl}
\end{itemize}

%% file: tex/Appendix/PatchExample.tex
\section{Code Examples}
\label{app:examples}

\input{figures/PatchedDiscrepancyExample}

\input{figures/UB_example}

%% file: figures/PatchedDiscrepancyExample.tex
\begin{figure}[htbp]
\centering
\begingroup
\definecolor{patchbeforebg}{RGB}{250,248,244}
\definecolor{patchafterbg}{RGB}{244,246,248}
\definecolor{hldel}{RGB}{145,105,20}
\definecolor{hladd}{RGB}{0,90,170}
\definecolor{cm}{RGB}{64,145,100}
\lstdefinestyle{patch-example}{
  style=code,
  columns=fullflexible, keepspaces=true,
  breakatwhitespace=true, breakindent=1.5em,
  commentstyle=\color{cm}\bfseries,
  xleftmargin=13pt, xrightmargin=5pt,
  framexleftmargin=10pt, framexrightmargin=2pt,
  aboveskip=0pt, belowskip=2pt,
}
\captionsetup{skip=5pt}
\captionsetup[subfigure]{skip=3pt}
\begin{subfigure}[t]{\linewidth}
\begin{lstlisting}[style=patch-example,backgroundcolor=\color{patchbeforebg}]
int fd = open("./empty_file", O_RDWR);
struct stat st;  fstat(fd, &st);
int size = st.st_size;  // == 0
«Dchar *p = (char*)mmap(NULL, size, PROT_READ, MAP_PRIVATE, fd, 0);D»
\end{lstlisting}
\caption{Unguarded \texttt{mmap} call.}
\label{fig:fix-before}
\end{subfigure}

\vspace{2pt}
\begin{subfigure}[t]{\linewidth}
\begin{lstlisting}[style=patch-example,backgroundcolor=\color{patchafterbg}]
int fd = open("./empty_file", O_RDWR);
struct stat st;  fstat(fd, &st);
int size = st.st_size;  // == 0
«Achar *p;
if (size == 0) { p = (char*)MAP_FAILED; errno = EINVAL; } 
else { p = (char*)mmap(NULL, size, PROT_READ, MAP_PRIVATE, fd, 0); }A»
\end{lstlisting}
\caption{Guarded \texttt{mmap} call for empty files.}
\label{fig:fix-after}
\end{subfigure}
\caption{Patch for the discrepancy in Figure~\ref{fig:mmap-bug}. The guard avoids the Emscripten crash for zero-length mapping requests. The original and replacement statements are highlighted in \textcolor{hldel}{\textbf{yellow}} and \textcolor{hladd}{\textbf{blue}}, respectively.}
\Description{Two vertically stacked C listings compare the original statement with its replacement. Both open an empty file and obtain its size with fstat. The upper listing assigns the result of an unconditional mmap call to pointer p. The lower listing first tests whether size equals zero. It calls mmap with the original arguments only when the mapping length is nonzero.}
\label{fig:mmap-fix}
\endgroup
\end{figure}

%% file: figures/UB_example.tex
\begin{figure}[h]
\centering
\begingroup
\definecolor{bg}{RGB}{244,246,248}
\definecolor{cm}{RGB}{64,145,100}
\definecolor{shellbg}{RGB}{250,248,244}
\lstdefinestyle{ub-code}{
  style=code,
  columns=fullflexible, keepspaces=true,
  breakatwhitespace=true,
  commentstyle=\color{cm}\bfseries,
  xleftmargin=13pt, xrightmargin=5pt,
  framexleftmargin=10pt, framexrightmargin=2pt,
}
\lstdefinestyle{ub-shell}{
  style=shell,
  columns=fullflexible, keepspaces=true,
  breakatwhitespace=true,
  basicstyle=\ttfamily\scriptsize\bfseries\color{black},
  moredelim=[is][\color{shellcmd}\mdseries]{`C}{C`},
}
\captionsetup{skip=5pt}
\captionsetup[subfigure]{skip=3pt}
\begin{subfigure}[h]{\linewidth}
\begin{lstlisting}[style=ub-code]
int first(void) { return 10; }
int second(void) { return 20; }
int func(int a, int b) { return b + a; }
int main() { // Execution order of `first` and `second` is unspecified.
    int result = func(first(), second());
    printf("Result: %d\n", result);
    return 0;
}
\end{lstlisting}
\label{fig:ub-code}
\end{subfigure}

\caption{Example of unspecified behavior: the order of evaluation of function arguments is unspecified by the C standard, so either \texttt{first()} or \texttt{second()} may be evaluated first.}
\Description{Functions first and second return 10 and 20, respectively. Both are passed as arguments to func, which adds its arguments. The order of evaluation of function arguments is unspecified by the C standard, so either \texttt{first()} or \texttt{second()} may be evaluated first.}
\label{fig:ub-bug}
\endgroup
\end{figure}

%% file: tex/Appendix/AddtionalExp.tex
\newpage
\input{tables/AppendixExperiments}
\section{Additional Experiments}
\label{appendix:additional_experiments}

\input{tex/Appendix/ManualFixAnalysis}

\input{tex/Appendix/Efficiency_details}

\input{tex/Appendix/AdditionalAblation}

%% file: tables/AppendixExperiments.tex
\begin{table}[h]
  \centering
  \small
  \caption{Manual review of fixes on the 40-case sample. Each model's fix is counted separately.}
  \label{tab:manual-fix-analysis}
  \begin{tabular}{lrrr}
    \toprule
    Model & Fixed & Good & Low Quality \\
    \midrule
    DeepSeek-Flash & 34 & 31 & 3 \\
    MiMo-Flash     & 24 & 22 & 2 \\
    MiniMax-M3     & 24 & 21 & 3 \\
    Muse-Glimmer  & 28 & 25 & 3 \\
    Qwen-Flash    & 30 & 26 & 4 \\
    \midrule
    \textbf{Total} & \textbf{140} & \textbf{125} & \textbf{15} \\
    \bottomrule
  \end{tabular}
\end{table}

\begin{table*}[!t]
\centering
\small
\caption{Average token usage and API cost (USD) per case across all 114 cases, including unsuccessful attempts.
New and Cached denote new and cached input tokens, and Out denotes output tokens (K = 1,000).
The cost of \tool includes Differential Trace Analysis (\$0.0018 per case).}
\label{tab:tool_efficiency_all_cases}
\setlength{\tabcolsep}{4pt}
\renewcommand{\arraystretch}{1.20}

\begin{tabular*}{0.98\textwidth}
    {@{}l@{\hspace{28pt}}
     *{2}{c@{\extracolsep{\fill}\hspace{4pt}}r@{\extracolsep{0pt}\hspace{18pt}}}
     c@{\extracolsep{\fill}\hspace{4pt}}r@{\extracolsep{0pt}\hspace{9pt}}
     |@{\hspace{9pt}}c@{\extracolsep{\fill}\hspace{4pt}}r@{\extracolsep{0pt}}}
\toprule
\multirow{2}{*}{Model}
& \multicolumn{2}{@{}c@{\hspace{18pt}}}{OpenCode}
& \multicolumn{2}{@{}c@{\hspace{18pt}}}{OpenHands}
& \multicolumn{2}{@{}c@{\hspace{\dimexpr18pt+\arrayrulewidth\relax}}}{SWE-agent}
& \multicolumn{2}{@{}c@{}}{\textbf{WasmMend}} \\
\cmidrule(r{18pt}){2-3}\cmidrule(r{18pt}){4-5}
\cmidrule(r{\dimexpr18pt+\arrayrulewidth\relax}){6-7}\cmidrule{8-9}
& New/Cached/Out & Cost
& New/Cached/Out & Cost
& New/Cached/Out & Cost
& New/Cached/Out & Cost \\
\midrule

DeepSeek-Flash
& 42K/1,998K/18K & 0.046
& 54K/1,463K/21K & 0.050
& 37K/1,857K/21K & 0.048
& 87K/952K/22K & 0.059 \\

MiMo-Flash
& 128K/4,702K/41K & 0.043
& 84K/1,039K/9K & 0.017
& 84K/2,435K/23K & 0.025
& 166K/2,166K/17K & 0.036 \\

MiniMax-M3
& 61K/1,692K/12K & 0.134
& 124K/1,821K/43K & 0.198
& 87K/2,011K/20K & 0.171
& 271K/1,906K/29K & 0.232 \\

Muse-Glimmer
& 88K/2,081K/11K & 0.131
& 109K/1,410K/17K & 0.120
& 67K/1,659K/16K & 0.114
& 163K/800K/13K & 0.111 \\

Qwen-Flash
& 77K/2,420K/13K & 0.056
& 110K/1,773K/14K & 0.051
& 90K/2,359K/16K & 0.059
& 123K/1,199K/11K & 0.045 \\

\bottomrule
\end{tabular*}
\end{table*}

\begingroup
\captionsetup{hypcap=false}
\begin{table}[!t]
\vspace{0pt}
  \centering
  \caption{Fix rates and costs per successful fix on the subset of 40 discrepancies.
    Costs are reported in USD.}
  \label{tab:opencode+trace}
  \small
  \setlength{\tabcolsep}{2.5pt}
  \begin{tabular}{@{}l rr rr rr@{}}
    \toprule
    \multirow{2}{*}{Model}
      & \multicolumn{2}{c}{OpenCode}
      & \multicolumn{2}{c}{\shortstack{OpenCode +\\Trace Analysis}}
      & \multicolumn{2}{c}{\shortstack{WasmMend\\(default)}} \\
    \cmidrule(lr){2-3}\cmidrule(lr){4-5}\cmidrule(l){6-7}
      & Fix Rate & Cost (\$)
      & Fix Rate & Cost (\$)
      & Fix Rate & Cost (\$) \\
    \midrule
    DeepSeek-Flash   & 55.0 & 0.034 & 62.5 & 0.040 & 85.0 & 0.050 \\
    MiMo-Flash       & 0.0 & N/A & 5.0 & 0.005 & 60.0 & 0.027 \\
    MiniMax-M3       & 27.5 & 0.087 & 42.5 & 0.139 & 60.0 & 0.102 \\
    Muse-Glimmer     & 52.5 & 0.091 & 60.0 & 0.085 & 70.0 & 0.094 \\
    Qwen-Flash       & 20.0 & 0.032 & 25.0 & 0.041 & 75.0 & 0.029 \\
    \midrule
    Average          & 31.0 & 0.061 & 39.0 & 0.062 & 70.0 & 0.060 \\
     \bottomrule
  \end{tabular}
\end{table}
\begin{table}[!t]
\vspace{0pt}
  \centering
  \caption{OpenCode fix rates (\%) with and without trace analysis,
    grouped by trace analysis quality, excluding \Mimo.
    $\Delta$ is the change in percentage points.}
  \label{tab:trace-quality-gains}
  \small
  \setlength{\tabcolsep}{12pt}
  \begin{tabular}{@{}l ccr@{}}
    \toprule
    Category & \shortstack{Plain \\OpenCode} & \shortstack{OpenCode + \\Trace Analysis} & $\Delta$ \\
    \midrule
    Precise           & 42.2 & 54.7 & $+12.5$ \\
    Report as Suspect & 17.9 & 10.7 & $-7.2$  \\
    Missing           & 75.0 & 75.0 & $0$     \\
    \midrule
    Overall           & 38.8 & 47.5 & $+8.8$  \\
    \bottomrule
  \end{tabular}
\end{table}
\endgroup

%% file: tex/Appendix/ManualFixAnalysis.tex
\subsection{Manual Analysis of Fixed Cases}
\label{appendix:manual_fix_analysis}

We manually review the fixes produced by \tool on the same subset of 40 annotated cases discussed in Section~\ref{ssec: trace_analysis_quality} across five backbone models. A fix is classified as good only if it makes Wasm match native behavior while preserving the program's original functionality and not impairing any other features; fixes that fail any criterion are classified as low quality. Table~\ref{tab:manual-fix-analysis} reports the results. Overall, 125 of the 140 reviewed fixes (89.3\%) are considered to be of good quality. Low-quality fixes typically silently alter functionality to pass the tests at the expense of security checks or native-build compatibility. 
Even after multiplying the fix rate reported in Table~\ref{tab:effectiveness} by the proportion of good fixes to estimate a quality-adjusted fix rate, \tool continues to substantially outperform the baselines.

%% file: tex/Appendix/Efficiency_details.tex
\subsection{Efficiency of \tool}
\label{appendix:full_efficiency}

In Table~\ref{tab:tool_efficiency_all_cases}, we provide the token usage and corresponding cost averaged over all 114 cases, including unsuccessful repairs. Across most backbone models, \tool's average cost per case over all cases is comparable to that of the open-source coding agents. The exception is \Minimax, with which \tool consumes substantially more new input tokens than with the other backbone models. Since new input tokens are more expensive than cached input tokens, this greater consumption raises \Minimax's average API cost above the baselines'.

%% file: tex/Appendix/AdditionalAblation.tex
\subsection{Additional Ablation Studies}
\label{appendix:addtional_ablation_studies}

On the 40 annotated discrepancies described in Section~\ref{ssec: trace_analysis_quality}, we compare OpenCode, OpenCode augmented with differential trace analysis, and the complete \tool system. OpenCode + Trace Analysis approximates a single-agent alternative to \tool without platform-aware design, using one agent to interpret trace analysis results and perform repairs.

Table~\ref{tab:opencode+trace} shows that trace analysis improves OpenCode's fix rate across all five models, raising the average from 31.0\% to 39.0\%. This improvement further demonstrates the benefits of differential trace analysis. The complete \tool system achieves 70.0\%, still substantially higher than OpenCode augmented with trace analysis. This additional gain supports the contribution of the platform-aware multi-agent repair framework to the effectiveness of the complete pipeline.

Table~\ref{tab:trace-quality-gains} compares OpenCode's fix rate with and without trace analysis across 160 repair runs with different models, grouped by the trace analysis quality discussed in Section~\ref{ssec: trace_analysis_quality}. We exclude \Mimo's 40 runs because it fixes only two cases even with trace guidance. The improvement is driven by \textit{Precise} cases (+12.5 percentage points), while fix rates decrease for \textit{Report as Suspect} cases and remain unchanged for \textit{Missing} cases. This suggests that trace guidance is most useful when it precisely localizes the root cause.

%% file: tex/Appendix/DatasetCategorization.tex
\newpage
\section{Dataset Categorization}
\label{appendix:dataset_categorization}

The full benchmark contains 114 discrepancy cases across 91 repositories (37 C and 54 C++), comprising 7.62 million lines of C/C++ code with a median of 19.9K lines per repository. These repositories contain 7,860 test files and 17,829 test cases, with medians of 30 files and 45 cases per repository, respectively.

We categorize the 40 manually reviewed discrepancies into root-cause categories adapted from WasmChecker~\cite{WasmChecker}, as shown in Table~\ref{tab:case-categories}.

\begin{table}[h]
  \centering
   \setlength{\tabcolsep}{11pt}
  \caption{Categorization of the 40 reviewed cases.}
  \label{tab:case-categories}
  \begin{tabular}{lr}
    \toprule
    Category & Cases \\
    \midrule
    Execution model & 16 \\
    Host environment \& memory model & 8 \\
    WebAssembly language features & 6 \\
    Compiler bugs \& standard library differences & 10 \\
    \midrule
    \textbf{Total} & \textbf{40} \\
    \bottomrule
  \end{tabular}
\end{table}

%% file: tex/Appendix/AdditionalInfoforMethodology.tex
\clearpage
\section{Algorithm for Wasm Event Matching}
\label{appendix:supp_of_diff_trace_analysis}

Algorithm~\ref{alg:match_wasm_event} details \texttt{MatchWasmEvent}, including candidate selection, state comparison, and tracking of previously matched Wasm events.

\begin{algorithm}[h]
\setstretch{0.9}%
\newcommand\mycommfont[1]{\textcolor{deepgreen}{#1}}
\SetCommentSty{mycommfont}
\DontPrintSemicolon
\caption{Wasm Event Matching (\texttt{MatchWasmEvent})}
\label{alg:match_wasm_event}
\footnotesize 

\KwIn{\texttt{evt}: The target native event to match \\
      \texttt{WasmEvents}: The indexed pool of Wasm events}
\KwOut{Boolean indicating if an equivalent Wasm event exists}
\BlankLine

\ForAll{\texttt{w\_evt} $\in$ \texttt{WasmEvents}}{
    \tcp{Skip events that have already been successfully matched}
    \If{\texttt{w\_evt.IsUsed}}{
        \textbf{continue}\;
    }
    \BlankLine
    \tcp{Require the same entry/exit kind}
    \If{\texttt{IsEntry(evt)} $\neq$ \texttt{IsEntry(w\_evt)}}{
        \textbf{continue}\;
    }
    \If{\texttt{evt.FuncID} $==$ \texttt{w\_evt.FuncID}}{
        \texttt{is\_equivalent} $\gets$ \text{True}\;
        \BlankLine
        
        \tcp{Iterate through detailed state variables}
        \ForAll{$(\texttt{n\_val}, \texttt{w\_val}) \in \texttt{Zip(evt.States, w\_evt.States)}$}{
            \uIf{\texttt{IsMemOrFD(n\_val)} $\land$ \texttt{IsMemOrFD(w\_val)}}{
                \tcp{Normalize platform-dependent values before comparison}
                \If{\texttt{Normalize(n\_val)} $\neq$ \texttt{Normalize(w\_val)}}{
                    \texttt{is\_equivalent} $\gets$ \text{False}\;
                    \textbf{break}\;
                }
            }
            \Else{
                \tcp{Direct comparison for standard values}
                \If{\texttt{n\_val} $\neq$ \texttt{w\_val}}{
                    \texttt{is\_equivalent} $\gets$ \text{False}\;
                    \textbf{break}\;
                }
            }
        }
        \BlankLine
        
        \If{\texttt{is\_equivalent}}{
            \texttt{w\_evt.IsUsed} $\gets$ \text{True} \tcp{Mark as used to prevent duplicate mapping}
            \Return \text{True} \tcp{Equivalent match found}
        }
    }
}
\BlankLine
\Return \text{False} \tcp{No matching event found in the pool}
\end{algorithm}